\documentclass[12pt]{iopart}

\usepackage{graphicx}
\begin{document}
\title[]{Effective field investigation of dynamic phase transitions for site diluted Ising ferromagnets driven by a periodically oscillating magnetic field}
\author{U. Akinci, Y. Yuksel, E. Vatansever and H. Polat$^\dagger$} 
\address{Deparment of Physics, Dokuz Eylul University, Buca/Izmir-35160, Turkey}
\ead{hamza.polat@deu.edu.tr}
\begin{abstract}
Dynamic behavior of a site diluted Ising ferromagnet in the presence of periodically oscillating magnetic field has been analyzed by
means of the effective field theory (EFT). Dynamic equation of motion have been solved for a honeycomb lattice ($z=3$) with the help of a
Glauber type stochastic process. The global phase diagrams and the variation of the corresponding dynamic order parameter as a function
of the Hamiltonian parameters and temperature has been  investigated in detail and it has been shown that the system exhibits reentrant
phenomena, as well as a dynamic tricritical point which disappears for sufficiently weak dilution.
\end{abstract}

\maketitle
\section{Introduction}
Ising model under a time dependent external magnetic field (kinetic Ising model) has attracted much interest from both theoretical and
experimental points of view recently. Due to the competing time scales of the system and periodic external magnetic field, the system cannot
respond to the external magnetic field instantaneously which causes interesting behaviors. At high temperatures for the high amplitudes of the
periodic magnetic field, the system follows the external field with a delay while this is not the case for low temperatures and small magnetic
field amplitudes. This spontaneous symmetry breaking indicates the presence of a dynamical phase transition (DPT)\cite{r1} which shows itself in
the  dynamical order parameter (DOP) which is defined as the time average of the magnetization over a full period of the oscillating field. DPT
is firstly observed theoretically within the mean field approximation (MFA) \cite{r2}. Since then, much attention has been devoted to investigate the dynamic nature of the phase transition in experimental studies. As an example, a DPT occurring in high frequency region is represented by Jiang et al. \cite{r3} using the surface magneto-optical Kerr-effect technique for epitaxially grown ultrathin Co films on a Cu(001) surface. For a [Co($4\mathrm{A^{o}}$)/Pt($7\mathrm{A^{o}}$)] multilayer system with strong perpendicular anisotropy, an example of DPT is shown by Robb et al.  \cite{r4}. In this study, they found that the experimental nonequilibrium phase diagrams are found to strongly resemble the dynamic behavior calculated from simulations of a kinetic Ising model. Thus, a strong evidence of consistency  between theoretical and experimental studies  is shown.  Polyethylene naphtalate nanocomposites are investigated by Kanuga et al. \cite{r5}. After detailed analysis, they shown that the material undergoes three critical structural transitions.

DPT and hysteresis behaviors of the S-1/2 Ising model with periodic external magnetic field has been studied by means of MFA \cite{r6,r7,r8,r9,r11}
and Monte Carlo (MC) simulations \cite{r12,r13,r14,r15,r16,r17,r18}, and there are also some works for higher spins \cite{r19,r20} or mixed spin
Ising systems \cite{r21,r22}, as well as the other variants of the model such as the transverse kinetic Ising model \cite{r23} based on the MFA.

It is well known that MFA neglects the thermal fluctuations via neglecting the self spin correlations whereas the effective field theory (EFT)
which is superior to conventional MFA is based on the differential operator technique \cite{rek_1,rek_2}  and it takes into account the self
spin correlations. Thus, it is expected from EFT to obtain more reasonable results than MFA for these systems, as in the case of static Ising model.
EFT has been applied to the kinetic Ising model in the past few years. For example, thermal and magnetic properties of kinetic Ising model \cite{r24,r25,r26,r27},
and also the transverse kinetic Ising model \cite{r28,r29} have been studied by EFT very recently. Based on the results obtained in these works,
we can conclude that EFT gives more reasonable results for the investigation of the DPT behaviors of the kinetic Ising model, in comparison to MFA.

On the other hand, as far as we know, the thermal and magnetic properties of the quenched disordered systems such as site (or bond) diluted
ferromagnets driven by a periodically oscillating magnetic field have not yet been investigated within the framework of EFT or MFA .
Furthermore, DPT properties of such disordered materials have not yet also been considered by means of MC simulations.
These types of disorder effects constitute an important role in material science,  since the quenched disorder effects may induce
some important macroscopic effects on the material. Therefore, we believe that the investigation of quenched randomness effects on
the DPT properties of Ising model still needs particular attention and we need a detailed consideration of
competition between the disorder conditions and the oscillation of magnetic field. Hence, in this work we intend to study the DPT
properties of a kinetic Ising system by introducing the quenched site dilution effects. For this purpose, we organized the paper
as follows: In Sec. \ref{formulation} we briefly present the formulations. The results and discussions are summarized in Sec. \ref{results},
and finally Sec. \ref{conclusion} contains our conclusions.

\section{Formulation}\label{formulation}
We consider an Ising ferromagnet defined on a lattice which has a coordination number $z$ with a time dependent external magnetic field.
The Hamiltonian describing our model is
\begin{equation}\label{denk1}
\mathcal{H}=-J\sum_{<i,j>}{}{c_ic_js_is_j}-H(t)\sum_{i}{c_is_i},
\end{equation}
where $J>0$ is spin-spin exchange interaction, $c_i$ is a site occupation variable and $s_i$ is the spin variable. Site occupation variable can
take the values $c_i=0$ which means that the site $i$ is empty  or $c_i=1$ if the site $i$ is occupied by a magnetic atom and the spin variable
can take values $s_i=\pm 1$.  The first summation in Eq. (\ref{denk1}) is over the nearest neighbor site pairs and the second one is over all
lattice sites. The time dependent external magnetic field is given by
\begin{equation}\label{denk2}
 H(t)=H_{0}\cos(\omega t),
\end{equation}
where $t$ is the time and  $H_{0}$ is the amplitude of the oscillating magnetic field with a frequency $\omega$.
The dynamical evolution of the  system may be given by Glauber dynamics\cite{r30} based on a master equation
\begin{equation}\label{denk3}
\tau\frac{d}{dt}\langle\langle c_{i}s_{i}\rangle\rangle_{r}=-\langle\langle c_{i}s_{i}\rangle\rangle_{r}+\langle\langle c_i\tanh \left[\beta c_i\left(E_i+H(t)\right)\right] \rangle\rangle_{r},
\end{equation}
where $1/\tau$ is the transition per unit time in a Glauber type stochastic process, $\beta=1/k_BT$ and $k_B$ represents the Boltzmann constant, $T$ is the temperature and $E_i$ is the local
field acting on the site $i$ and it is given by
\begin{equation}\label{denk4}
E_i=J\sum_{\delta=1}^{z}{c_\delta s_\delta},
\end{equation}
where $z$ is the coordination number of the lattice. The inner average brackets in Eq. (\ref{denk3}) stands for the thermal average and the outer
one (which has a subscript $r$) represents the random configurational average which is necessary for including the site dilution effects.

In order to handle the second term on the right hand side of the Eq. (\ref{denk3}) one can use the differential operator technique
\cite{rek_1,rek_2}. By using the differential operator technique, Eq. (\ref{denk3}) gets the form
\begin{equation}\label{denk5}
\tau\frac{d}{dt}\langle\langle c_{i}s_{i}\rangle\rangle_{r}=-\langle\langle c_{i}s_{i}\rangle\rangle_{r}+\langle\langle c_i\exp (c_iE_i\nabla) \rangle\rangle_{r} f(x),
\end{equation}
where $\nabla=\partial/\partial x$ is the differential operator and the function $f(x)$ is given by
\begin{equation}\label{denk6}
f(x)=\tanh\left[\beta( x+H(t))\right].
\end{equation}
The effect of the differential operator on a function $f(x)$ is given by
\begin{equation}\label{denk7}
\exp{(a\nabla)}f(x)=f(x+a),
\end{equation}
with any constant  $a$. By taking into account the two possible values of $c_i$ as $c_i=0,1$ we can write the exponential term as
\begin{equation}\label{denk8}
\exp{(ac_i)}=c_i\exp{(a)}+1-c_i,
\end{equation}
where $a$ is any constant. By using Eq. (\ref{denk8}) in Eq. (\ref{denk5}) we can obtain
\begin{equation}\label{denk9}
\tau\frac{d}{dt}\langle\langle c_{i}s_{i}\rangle\rangle_{r}=-\langle\langle c_{i}s_{i}\rangle\rangle_{r}+\langle\langle c_i\exp (E_i\nabla)\rangle\rangle_{r} f(x),
\end{equation}
where for site occupations $c_i^2=c_i$  was used. By using Eq. (\ref{denk4}) in Eq. (\ref{denk9}) we get
\begin{equation}\label{denk10}
\tau\frac{d}{dt}\langle\langle c_{i}s_{i}\rangle\rangle_{r}=-\langle\langle c_{i}s_{i}\rangle\rangle_{r}+\left\langle \left \langle c_{i}\prod_{\delta=1}^{z}\left[c_\delta\exp(J s_\delta \nabla)+1-c_\delta\right] \right\rangle\right\rangle_{r}f(x).
\end{equation}
In order to get a polynomial form of the second term on the right hand side of Eq. (\ref{denk10}), we write the exponential term in terms
of the hyperbolic trigonometric functions
\begin{equation}\label{denk11}
\begin{array}{lcl}
&&\displaystyle\tau\frac{d}{dt}\langle\langle c_{i}s_{i}\rangle\rangle_{r}=-\langle\langle c_{i}s_{i}\rangle\rangle_{r}\\
&&\displaystyle+\left\langle \left \langle c_{i}\prod_{\delta=1}^{z}\left[c_\delta\cosh(J\nabla)+c_\delta s_\delta\sinh(J \nabla)+ 1-c_\delta\right] \right\rangle\right\rangle_{r}f(x).
\end{array}
\end{equation}
When the product in Eq. (\ref{denk11}) is expanded, the multi site spin correlations appear. For simplicity let us handle these correlations with an improved decoupling
approximation (DA)\cite{r32} as
\begin{equation}\label{denk12}
\langle\langle c_i\ldots c_j c_ks_k\ldots c_ls_l \rangle\rangle_{r}=\langle\langle c_{i} \rangle\rangle_{r}...\langle\langle c_{j} \rangle\rangle_{r}\langle\langle c_{k}s_{k} \rangle\rangle_{r}...\langle\langle c_{l}s_{l} \rangle\rangle_{r},
\end{equation}
with
\begin{equation}\label{denk13}
\langle c_{i} \rangle_{r}=c, \quad \quad \langle\langle c_{i}s_{i} \rangle\rangle_{r}=m.
\end{equation}
Detailed discussion of DA in site dilution problem for the static case can be found in\cite{r32}.
By using Eqs. (\ref{denk12}) and (\ref{denk13}) in Eq. \ref{denk11} we get
\begin{equation}\label{denk14}
\tau\frac{dm}{dt}=-m+c\left\langle \left \langle\left[c \cosh(J\nabla)+m\sinh(J\nabla)+1-c\right]^{z}  \right\rangle\right\rangle_{r}f(x),
\end{equation}
By using the binomial expansion and writing the hyperbolic trigonometric functions in terms of the exponential functions
we get the compact form of Eq. (\ref{denk14}) as
\begin{equation}\label{denk15}
\frac{dm}{dt}=\frac{1}{\tau}\left( -m+\sum_{r=0}^{z}A_{r}m_{r}\right),
\end{equation}
where
\begin{equation}\label{denk16}
A_{r}=\sum_{p=r}^{z}\left ( \begin{array}{c}
                              z \\
                              p
                            \end{array}
\right)
\left ( \begin{array}{c}
                              p \\
                              r
                            \end{array}
\right)c^{p-r+1}(1-c)^{z-p}\cosh ^{p-r}(J \nabla)\sinh ^{r}(J \nabla)f(x).
\end{equation}
Eq. (\ref{denk15}) can be solved by various numerical methods. In this work we prefer to use the fourth order Runge-Kutta method (RK4) by
regarding Eq. (\ref{denk15}) as an initial value problem. We can mention that the differential equation derived in Eq. (\ref{denk15}) extends
up to the term  $m^{z}$. Each term in Eq. (\ref{denk15}) makes contribution to the solution, because the value of $m$, which is calculated
at each time step, is related to the previous $m$ value however the situation is different from the behavior of the equilibrium systems
at which high ordered terms can be neglected in the neighborhood of phase transition point.

The system has three Hamiltonian parameters, namely magnetic field frequency ($\omega$), magnetic field amplitude ($H_0$) and site
concentration ($c$). For selected values of these Hamiltonian parameters and temperature, RK4 will give convergency behavior after some
iterations i.e. the solutions have property $m(t)=m(t+2\pi/\omega)$ for arbitrary initial value for the magnetization
($m(t=0)$). Thus, after obtaining this convergent region after some transient steps (which depends on Hamiltonian parameters
and the temperature) the DOP can be calculated from
\begin{equation}\label{denk17}
Q=\frac{\omega}{2\pi}\oint m(t)dt.
\end{equation}

The dynamic nature of the system for a given temperature and Hamiltonian parameters can be determined from the DOP.
There are three possible states for the system, namely ferromagnetic (F), paramagnetic (P) and mixed (F+P). The solution $m(t)$ in the
convergent region satisfies
\begin{equation}\label{denk18}
m(t)=-m( t+\pi/\omega),
\end{equation}
in the P phase which is called the symmetric solution. The solution
corresponding to P phase follows the external magnetic field and oscillates around zero value which means that the DOP $Q$ is zero.
In the F phase, the solution does not satisfy Eq. (\ref{denk18}) and this solution is called as non-symmetric
solution which oscillates around a non-zero magnetization value, and does not follow the external magnetic field i.e. the value of
$Q$ is different from zero. In these two cases, the observed behavior of the magnetization is regardless of the choice of the initial
value of magnetization $m(0)$ whereas the last phase has magnetization solutions symmetric or non-symmetric depending on the  choice
of the initial value of magnetization corresponding to the coexistence region where F and P phases overlap.

\section{Results and Discussion}\label{results}

\begin{figure}\label{sek1}
\center
\includegraphics[width=10cm]{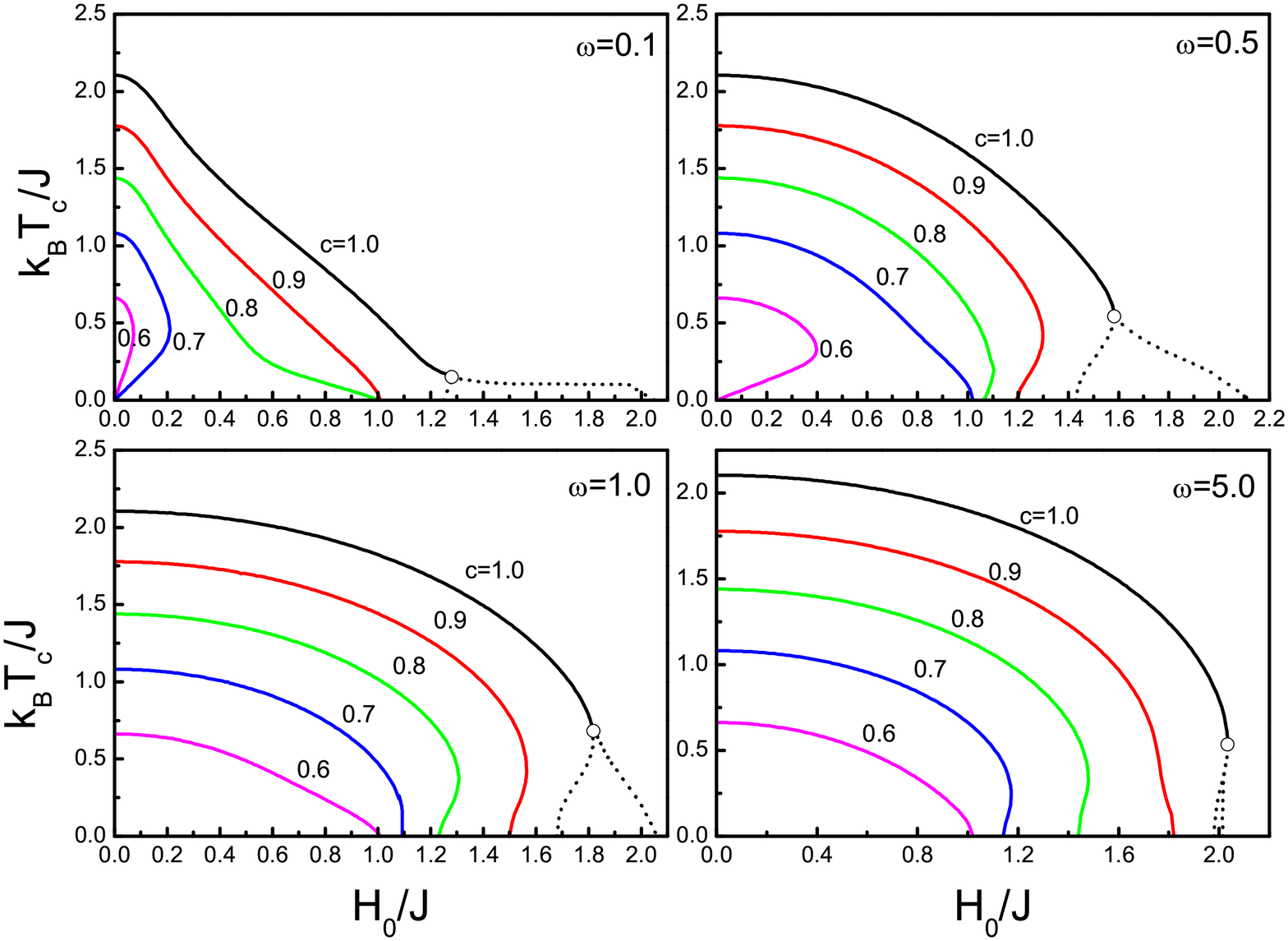}\\
\caption{Phase diagrams in $(k_{B}T_{c}/J-H_{0}/J)$ plane for a honeycomb lattice with some selected values of $\omega$ and $c$. The solid (dotted) lines correspond to the second (first) order transitions and solid symbols denote the dynamic tricritical points.}
\end{figure}

\begin{figure}\label{sek3}
\center
\includegraphics[width=10cm]{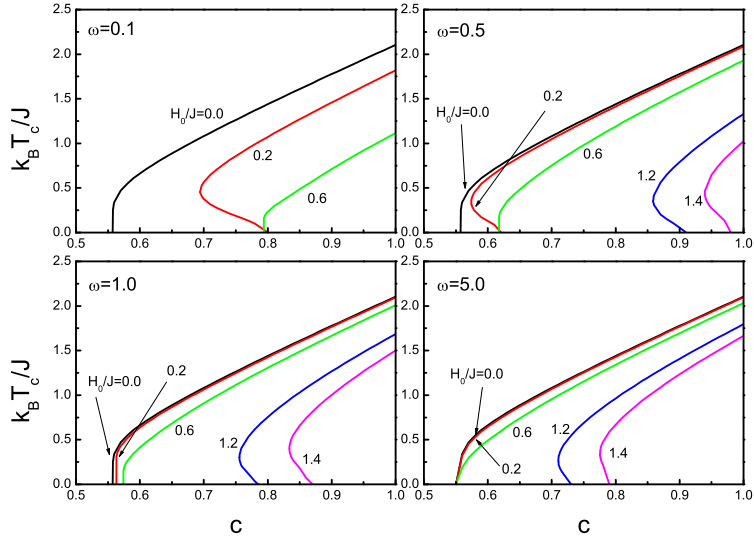}\\
\caption{Phase diagrams for a honeycomb lattice in $(k_{B}T_{c}/J-c)$ plane for some selected values of $\omega$ and $H_{0}/J$. The numbers accompanying each curve denote the value of the amplitude $H_{0}/J$ of the oscillating magnetic field.}
\end{figure}

In order to determine the effect of the Hamiltonian parameters on the critical temperature of the system we present the phase diagrams for a
honeycomb lattice $(z=3)$ in $(k_{B}T_{c}/J-H_{0}/J)$  and $(k_{B}T_{c}/J-c)$ planes with frequencies $\omega=0.1,0.5,1.0,5.0$  in Figs. (\ref{sek1})
and (\ref{sek3}), respectively.

At first sight we can see from Fig. (\ref{sek1}) that as $c$ value decreases then the $F$ phase region gets narrower in the $(k_{B}T_{c}/J-H_{0}/J)$ plane.
This is an expected result, since as the lattice sites are diluted  more and more then the energy contribution which comes from the spin-spin interactions gets
smaller. Hence, the system can undergo a DPT at smaller critical temperatures, due to the energy originating from the temperature and (or)
magnetic field which overcomes the spin-spin interactions. Therefore, the ferromagnetic regions in  $(k_{B}T_{c}/J-H_{0}/J)$ plane get narrower. In addition,
increasing $H_{0}/J$ values reduce the dynamical critical temperature while as $\omega$ increases then  the $F$ phase region gets wider in the
$(k_{B}T_{c}/J-H_{0}/J)$ plane for all concentration values. These observations can also be explained by the well known mechanisms underlying the
DPT's. In other words, increasing field amplitude facilitates the phase transition due to the increasing energy coming from
the oscillating magnetic field which tends to align the spins in its direction whereas an increasing field frequency gives rise to a growing
phase delay between the magnetization and magnetic field (i.e. the magnetization cannot follow the oscillating magnetic field) and this makes
the occurrence of the DPT difficult, as a result of this mechanism critical temperature increases. Moreover, the low concentration
phase diagrams in Fig. (\ref{sek1}), which appear at small $\omega$ values such as $w=0.1$ and $0.5$, exhibit a reentrant behavior and the reentrance
disappears at a certain value of increasing frequency between $0.5<\omega<1.0$. For example, as seen from the curve with $H_{0}/J=0.2, \omega=0.5$ and $c=0.6$,
corresponding to the upper right panel Fig. (\ref{sek1}), as the temperature increases starting from a value in the paramagnetic phase then the system undergoes
a second order phase transition  due to the thermal agitations and remains at a ferromagnetic order for a while. As the temperature  increases further then
the system undergoes another second order phase transition from a ferromagnetic to a paramagnetic phase.

\begin{figure}\label{sek2}
\center
\includegraphics[width=10cm]{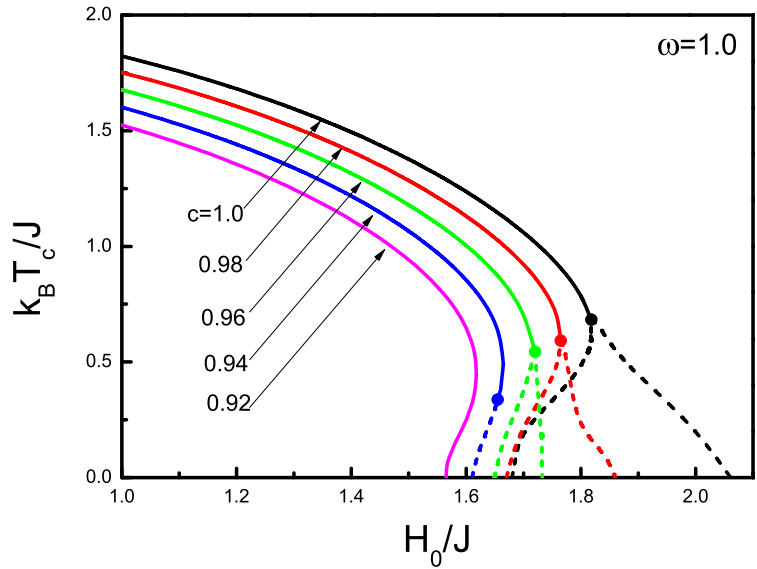}\\
\caption{Phase diagrams for a honeycomb lattice in ($k_{B}T_{c}/J-H_{0}/J)$ plane for $\omega=1.0$. The numbers accompanying each curve denote the value of site concentration $c$. The solid (dashed) curves correspond to the second (first) order transitions and solid symbols denote the dynamic tricritical points.}
\end{figure}

An interesting result is that the $F+P$ region disappears in the $(k_{B}T_{c}/J-H_{0}/J)$ plane for sufficiently weak dilution. In order to focus on this
interesting behavior, we depict the phase diagrams in the same plane for $\omega=1.0$ with the concentration values $c=1.00, 0.98, 0.96, 0.94,0.92$ in
Fig. (\ref{sek2}). As seen in Fig. (\ref{sek2}), as the concentration value decreases then $F+P$ region which appears at low temperatures and high field amplitude
values, gradually shrinks then disappears at a certain concentration value. Then we can not see this region in the phase diagrams which are shown in
Fig. (\ref{sek1}) except the concentration value $c=1.0$. On the other hand, one observes a dynamic tricritical point only in a narrow range of site
concentration $c$ which disappears at a certain concentration value. Namely, as seen in Fig. (\ref{sek2}) for $c=0.94$ we observe a reentrant behavior
of first order instead of the coexistence region while for $c=0.92$ the first order reentrance turns into a second order reentrance.

In order to probe the concentration dependence of the dynamic nature of the system, we plot the phase diagrams in $(k_{B}T_{c}/J-c)$ plane for some
selected values of Hamiltonian parameters in Fig. (\ref{sek3}). We can see that the results shown in Fig. (\ref{sek3}) are completely consistent with the
results presented in Fig. (\ref{sek1}). As the concentration $c$ of magnetic atoms  decreases then the ferromagnetic region gets narrower, as expected.
In Fig. (\ref{sek3}), we can also observe that there are mainly two reentrance regions which are located for $H_{0}/J=0.2$ with $\omega=0.1,0.5$ and
for $H_{0}/J=1.2,1.4$ with $\omega=0.5,1.0,5.0$, respectively. The former is nothing but just the reentrant region shown in the upper panels in
Fig. (\ref{sek1}) that observed for low concentration values. By taking into account these observations, we can conclude that the system may exhibit
reentrant behavior for a wide range of Hamiltonian parameters as a result of a competition between the interactions that exist in the system. Hence,
in the paramagnetic phase, the increasing temperature inhibits the ability of response of the magnetization to the external field even at low frequency
region by allowing a transition from paramagnetic phase to ferromagnetic one.

\begin{figure}\label{sek4}
\center
\includegraphics[width=10cm]{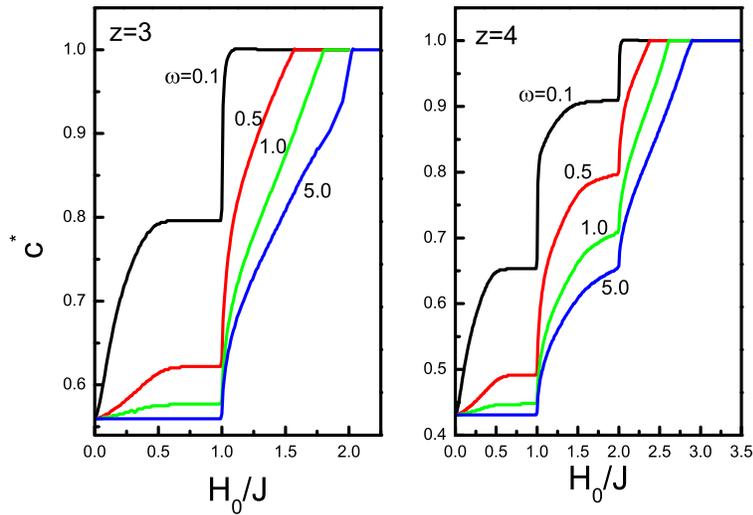}\\
\caption{Variation of the site percolation threshold $c^\star$ of honeycomb and square lattices with the amplitude $H_{0}/J$ of the oscillating magnetic field. The numbers accompanying each curve represent the frequency $\omega$ of the oscillating field.}
\end{figure}

At a certain $c$ value the system undergoes a phase transition at zero temperature which means that the system cannot exhibit an ordered phase
below this certain $c$ value for the phase diagrams in $(k_{B}T_{c}/J-c)$ plane.  No matter the phase diagrams in $(k_BT_c/J-c)$ plane shows
reentrance or not, for concentration values lower than this certain concentration value, namely site percolation threshold value ($c^\star$),
the system cannot exhibit an ordered phase at all. In other words, for $c>c^\star$ the system forms an infinite cluster of lattice sites.
However, as $c$ gets closer to $c^\star$ then isolated finite clusters appear, since the spin-spin interaction energy become insufficient to
compete with the magnetic field energy, even in the absence of the thermal energy (i.e. at zero temperature). As a result of this, the system
cannot exhibit long range ferromagnetic order even at zero temperature for $c<c^\star$. The variation of $c^\star$ as a function of the magnetic
field amplitude for frequencies $\omega=0.1,0.5,1.0$ and $5.0$ can be seen in Fig. (\ref{sek4}). As seen in Fig. (\ref{sek4}), $H_{0}/J$ and $\omega$
parameters have an opposite effect on $c^\star$. We can say that the general effect of increasing $H_{0}/J$ is to increase the value of $c^\star$
whereas increasing value of $\omega$ parameter decreases the value of $c^\star$. An important observation in Fig. (\ref{sek4}) is the appearance
of plateaus for different $\omega$ values. For $\omega=0.1$, $c^\star$ value is observed between $0.5<H_{0}/J<1.0$  which almost does not
change its shape with varying $H_{0}/J$ values. As $\omega$ value gets higher (e.g $\omega=5.0$ curve on left panel in Fig. (\ref{sek4}))
then another wider plateau originates within the range $0.0<H_{0}/J<1.0$. We can see from the right panel in Fig. (\ref{sek4}) that the
same situation holds for a square lattice $(z=4)$. We should state that in the absence of magnetic field $(H_{0}/J=0)$, $c^\star$ values
are exactly the same values with the static case obtained by the same approximation (DA) \cite{r33}.

\begin{figure}\label{sek6}
\center
\includegraphics[width=10cm]{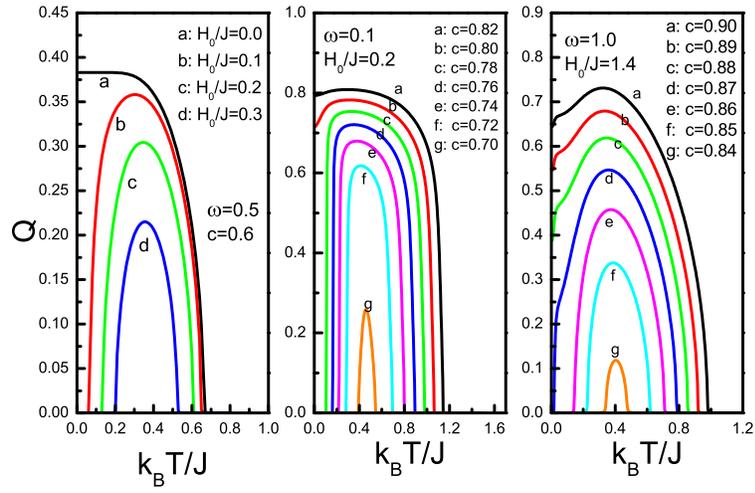}\\
\caption{Variation of the DOP $Q$ for a honeycomb lattice as a function of the temperature for some selected values of the Hamiltonian parameters $\omega, H_0/J$ and $c$.}
\end{figure}

It is well known that one cannot write an appropriate expression for the free energy including the presence of the time dependent
oscillating magnetic field. Hence, we cannot use the free energy expression to determine the type of the phase transition (first or second order).
Hence, we simply follow a procedure described briefly below. Since the time averaged magnetization over a full cycle of the external magnetic
field acts as the DOP then we may check the temperature dependence of it. Strictly speaking, if the DOP
decreases continuously to zero in the vicinity of critical temperature, this transition is classified as of the second order whereas if it vanishes
discontinuously then the transition is assumed to be of the first order. The results of the previously published works have pointed out that all
transitions from $F$ or $P$ phases to $F+P$ phase (or vice versa) are of the first order transitions \cite{r24,r25}. Hence we shall only investigate
the transitions between $F$ and $P$ phases, especially for the Hamiltonian parameters which exhibit reentrance. The variation of the DOP with the temperature for some selected values of the Hamiltonian parameters can be seen in Fig. (\ref{sek6}). From the left panel in
Fig. (\ref{sek6}), we can see that all transitions on the $c=0.6$ phase diagram with $\omega=0.5$ corresponding to the case in the upper right
panel in Fig. (\ref{sek1}) are of the second order. The situation is similar for the middle panel of Fig. (\ref{sek6}) where $H_0/J=0.2$ and $\omega=0.1$
corresponding to the upper right panel in Fig. (\ref{sek3}). Finally, we can conclude with the transitions shown in the phase diagrams
with $\omega=1.0$ and $H_0/J=1.4$ which are of the second order as seen in the right panel of Fig. (\ref{sek6}). A detailed investigation
for the variation of the DOP with the temperature for other Hamiltonian parameters shows us that all phase transitions
of the site diluted system are of the second order except the transitions from $F$ or $P$ phases to $F+P$ phase which occurs at low temperature
and high field amplitude values and within a very narrow region of the Hamiltonian parameter values (e.g. for $\omega=1.0, c=0.94$ curve as
shown in Fig.(\ref{sek2})).

\section{Conclusion}\label{conclusion}
In this work, we have investigated the dynamic nature of the critical phenomena which is observed for a site diluted Ising ferromagnet defined
on a honeycomb lattice $(z=3)$ driven by an external oscillating magnetic field by means of EFT. We have used a Glauber-type stochastic process
to describe the time evolution of the system. We have given the global phase diagrams, including the reentrant phase transitions. According to
our calculations, F+P phase disappears for sufficiently weak dilution of lattice sites.

EFT method takes the standard mean field predictions one step forward by taking into account the single site correlations which means that the thermal
fluctuations are partially considered within the framework of EFT. Although all of the observations reported in this work shows that EFT can be successfully
applied to such nonequilibrium systems in the presence of quenched site disorder, the true nature of the physical facts underlying the observations
displayed in the present work (especially the origin of the coexistence phase) may be further understood with an improved version of the present EFT
formalism which can be achieved by attempting to consider the multi site correlations which originate when expanding the spin identities.
We believe that this attempt could provide a treatment beyond the present approximation.

In conclusion, we hope that the results obtained in this work would shed light on the further investigations of the dynamic nature of the critical
phenomena in disordered systems and would be beneficial from both theoretical and experimental points of view.

\newpage

\newpage

\newpage

 \newpage

\newpage

\end{document}